\documentstyle[12pt]{article}
\newlength{\extraspace}
\newlength{\extraspaces}
\catcode`\@=11
\newcommand{\be}{\begin{equation}
\addtolength{\abovedisplayskip}{\extraspaces}
\addtolength{\belowdisplayskip}{\extraspaces}
\addtolength{\abovedisplayshortskip}{\extraspace}
\addtolength{\belowdisplayshortskip}{\extraspace}}
\newcommand{\ee}{\end{equation}}
\newcommand{\ba}{\begin{eqnarray}
\addtolength{\abovedisplayskip}{\extraspaces}
\addtolength{\belowdisplayskip}{\extraspaces}
\addtolength{\abovedisplayshortskip}{\extraspace}
\addtolength{\belowdisplayshortskip}{\extraspace}}
\newcommand{\ea}{\end{eqnarray}}
\newcommand{\nonu}{\nonumber \\[.5mm]}
\newcommand{\A}{&\!\!\!}

\newcommand{\tr}{\, {\rm tr}}
\newcommand{\e}{\, {\rm e}}
\newcommand{\D}{{\cal D}}
\newcommand{\bra}[1]{\left\langle {#1} \right\vert}
\newcommand{\ket}[1]{\left\vert {#1} \right\rangle}
\newcommand{\VEV}[1]{\left\langle {#1} \right\rangle}

\begin{document}
\addtolength{\baselineskip}{.7mm}
\thispagestyle{empty}
\begin{flushright}
STUPP--96--142 \\
{\tt hep-th/9602141} \\ 
February, 1996
\end{flushright}
\vspace{10mm}
\begin{center}
{\Large{\bf 
Ward-Takahashi Identities \\[2mm]
in Large $N$ Field Theories 
}} \\[25mm] 
{\sc Minako Araki}\footnote{
\tt e-mail: minako@th.phy.saitama-u.ac.jp} 
\hspace{3mm} and \hspace{3mm}
{\sc Yoshiaki Tanii}\footnote{
\tt e-mail: tanii@th.phy.saitama-u.ac.jp} \\[12mm]
{\it Physics Department, Faculty of Science \\[2mm]
Saitama University, Urawa, Saitama 338, Japan} \\[25mm]
{\bf Abstract}\\[7mm]
{\parbox{13cm}{\hspace{5mm}
The Ward-Takahashi identities in large $N$ field theories 
are expressed in a simple form using master fields. 
Operators appearing in these expressions are found to be 
generators of symmetry transformations acting 
on the master fields. 
}}
\end{center}
\vfill
\newpage
%
%%%%%  Introduction  %%%%%%%%%%%%%%%%%%%%%%%%%%%%%%%%%%%%%%%%%%%%%%
%
The technique of large $N$ limit is useful to study 
non-perturbative properties of quantum field theories. 
(For a review and references, see \cite{COL} for example.) 
When dynamical variables are $N \times N$ matrices, 
only planar Feynman diagrams contribute to the large $N$ 
limit \cite{THOOFT}. Such matrix models in the large $N$ limit 
were used to study SU($N$) gauge theories and 
two-dimensional quantum gravity. 
In studying the large $N$ field theories the idea of master 
fields was introduced \cite{WITTEN}. 
In the large $N$ limit path integrals are dominated by a single 
classical configuration called the master field, and 
all Green's functions are determined by it. 
\par
Recently, a mathematically precise formulation of master fields 
was given \cite{DOUGLAS}, \cite{GG} using non-commutative 
variables \cite{VDN}, and was further developed 
\cite{LI}--\cite{AV}. Similar formulations had previously been 
discussed in refs.\ \cite{HAAN}, \cite{CLS}. In this formulation 
master fields are given in terms of all connected Green's 
functions. One needs to solve the theories and obtain 
all Green's functions before constructing the master fields. 
At present it is not known how to construct master fields explicitly 
except for a few simple theories such as two-dimensional QCD 
and zero-dimensional matrix models. 
It remains to be found a method to construct master fields 
in more general theories. 
\par
On the other hand, one may use master fields to discuss formal 
properties of the theories. 
The purpose of this paper is to discuss the Ward-Takahashi identities 
corresponding to symmetry transformations in the large $N$ field 
theories. We find that the Ward-Takahashi identities can be expressed 
in a simple form using master fields. 
We also find that operators appearing in these expressions are 
generators of the symmetry transformations acting 
on the master fields. 
\par
%
%%%%%  master fields in large $N$ field theories  %%%%%%%%%%%%%%%%%
%
Let us first recall the method of master fields in large $N$ field 
theories \cite{DOUGLAS}, \cite{GG}, \cite{HAAN}, \cite{CLS}. 
We follow the approach in ref.\ \cite{CLS}. 
We shall consider general SU($N$) invariant field theories in which 
fields $\Phi_I$ take their values in $N \times N$ hermitian matrices. 
The index $I$ represents a type of the fields as well as space-time 
coordinates $x^\mu$, on which the fields depend. The fields can be 
either bosonic ($(-1)^{|I|} = + 1$) or fermionic ($(-1)^{|I|} = - 1$). 
The path integral has a form 
\be
\int \D \Phi_I \e^{i N S[\Phi]}, 
\ee
where coupling constants in the action $S[\Phi]$ are independent of $N$. 
We are interested in the large $N$ limit of vacuum expectation values 
\be
Z_{I_k \cdots I_1} = \lim_{N \rightarrow \infty} {1\over N} 
\bra{0} {\rm T} \tr ( \Phi_{I_k} \cdots \Phi_{I_1} ) \ket{0}. 
\label{vev}
\ee
In the perturbation theory they are given by a sum of planar 
Feynman diagrams. 
\par
As discussed in ref.\ \cite{CLS} it is convenient to introduce 
a non-commuting source $j_I$ for each field $\Phi_I$ and consider 
the generating functional 
\be
Z[j] = 1 + \sum_{k=1}^\infty \sum_{\{ I \}} 
j_{I_1} \cdots j_{I_k} Z_{I_k \cdots I_1}. 
\ee
If we define the differentiation with respect to the 
non-commuting sources by 
\be
{\delta \over \delta j_I} \, ( a j_{I_1} j_{I_2} \cdots j_{I_k} ) 
= a \, \delta_{I I_1} \, j_{I_2} \cdots j_{I_k}, \qquad 
{\delta \over \delta j_I} \, a = 0, 
\ee
where $a$ is a constant, 
the planar Green's functions (\ref{vev}) are given by 
\be
Z_{I_k \cdots I_1} = \left. {\delta \over \delta j_{I_k}} \cdots 
{\delta \over \delta j_{I_1}} \, Z[j] \right|_{j=0}. 
\label{vev2}
\ee
Similarly, the generating functional of connected planar Green's 
functions is defined by using non-commuting sources $J_I$ as 
\be
W[J] = \sum_{k=1}^\infty \sum_{\{ I \}} 
J_{I_1} \cdots J_{I_k} \lim_{N \rightarrow \infty} 
{1\over N} \bra{0} {\rm T} \tr ( \Phi_{I_k} \cdots \Phi_{I_1} ) 
\ket{0}_{\rm connected}. 
\ee
In ref.\ \cite{CLS} it was shown that these two generating 
functionals are related by 
\be
Z[j] = 1 + W[J], \qquad J_I = j_I Z[j]. 
\ee
\par
We define operators $\hat\Phi_I$ and $\hat\Pi_I$ by a similarity 
transformation 
\ba
\hat\Phi_I \A = \A Z[j]^{-1} {\delta \over \delta j_I} Z[j] 
= {\delta \over \delta J_I} 
+ {\delta W[J] \over \delta J_I}, \nonu
\hat\Pi_I \A = \A Z[j]^{-1} j_I Z[j] = (1 + W[J] )^{-1} J_I, 
\label{sim}
\ea
which satisfy 
\be
\hat\Phi_I \hat\Pi_J = \delta_{IJ}. 
\ee
In terms of these operators the planar Green's functions 
(\ref{vev2}) can be expressed as 
\ba
Z_{I_k \cdots I_1} 
\A = \A \left. \hat\Phi_{I_k} \cdots \hat\Phi_{I_1} \cdot 1 
\right|_{J=0} \nonu 
\A = \A \bra{\Omega} \hat\Phi_{I_k} \cdots 
\hat\Phi_{I_1} \ket{\Omega}, 
\ea
where the states $\ket{\Omega}$ and $\bra{\Omega}$ are defined by 
\be
{\delta \over \delta J_I} \ket{\Omega} = 0, \qquad 
\bra{\Omega} J_I = 0, \qquad \VEV{\Omega | \Omega} = 1. 
\ee
$\hat\Phi_I$ are the master field operators discussed in the recent 
works \cite{DOUGLAS}, \cite{GG}. The operators $J_I$ and 
${\delta \over \delta J_I}$ correspond to $\hat a_I^\dagger$ and 
$\hat a_I$ respectively in the notation of 
refs.\ \cite{DOUGLAS}, \cite{GG}. 
We note that these operators are not quantum operators but just 
represent $N \rightarrow \infty$ limit of $N \times N$ matrices. 
\par
%
%%%%%  Ward-Takahashi identities  %%%%%%%%%%%%%%%%%%%%%%%%%%%%%%%%%%%
%
We now turn to a discussion of the Ward-Takahashi identities. 
Suppose that there are $n$ conserved charges $Q^a$ 
$(a = 1, 2, \cdots, n)$ corresponding to symmetry transformations 
of the theory. These charges generate the transformations of the 
quantum fields $\Phi_I$ 
\be
\delta \Phi_I = 
[ i \lambda^a Q^a, \Phi_I ] \equiv \lambda^a R^a_I(\Phi), 
\label{generator}
\ee
where $\lambda^a$ are parameters of the transformations and have 
the same statistics as $Q^a$. 
They satisfy a commutation relation 
\be
[ Q^a, Q^b \} = i f^{abc} Q^c, 
\label{algebra}
\ee
where $[ A, B \}$ represents an anticommutator of $A$ and $B$ 
when both of $A$ and $B$ are fermionic, and a commutator 
otherwise. We assume that the structure constant $f^{abc}$ is 
bosonic. From the Jacobi identity 
\be
[ Q^a, [ Q^b, \Phi_I \} \} 
- (-1)^{|a||b|} [ Q^b, [ Q^a, \Phi_I \} \} 
= [ [ Q^a, Q^b \}, \Phi_I \}, 
\label{jacobi}
\ee
we obtain a condition on $R^a_I$ 
\be
[ Q^a, R^b_I \} - (-1)^{|a||b|} [ Q^b, R^a_I \} 
= i f^{abc} R^c_I. 
\label{jacobi2}
\ee
We will use this relation later. 
When the vacuum is invariant under the symmetry transformations 
$Q^a \ket{0} = 0$, we obtain the Ward-Takahashi identities 
\ba
0 \A = \A \bra{0} [ i \lambda^a Q^a, {\rm T} \tr ( \Phi_{I_k} 
\cdots \Phi_{I_1} ) ] \ket{0} \nonu 
\A = \A \sum_{i=1}^k \bra{0} {\rm T} \tr 
( \Phi_{I_k} \cdots \Phi_{I_{i+1}} \lambda^a R^a_{I_i} (\Phi) 
\Phi_{I_{i-1}} \cdots \Phi_{I_1} ) \ket{0}. 
\label{ward}
\ea
These identities are satisfied for an arbitrary $N$. 
\par
We would like to express these identities using the master fields 
$\hat\Phi_I$ in the large $N$ limit. By taking the limit 
$N \rightarrow \infty$ and 
multiplying non-commuting sources we obtain 
\be
\sum_{k=1}^\infty \sum_{\{I\}} j_{I_1} \cdots j_{I_k} 
\sum_{i=1}^k \lim_{N \rightarrow \infty} {1 \over N} 
\bra{0} {\rm T} \tr ( \Phi_{I_k} \cdots \Phi_{I_{i+1}} 
\lambda^a R^a_{I_i} (\Phi) \Phi_{I_{i-1}} \cdots \Phi_{I_1} ) 
\ket{0} = 0. 
\label{ward2}
\ee
To proceed, we expand the functions $R^a_I(\Phi)$ as 
\be
R^a_I(\Phi) = R^{(0)a}_I + \sum_K R^{(1)a}_{I, K} \Phi_K 
+ \sum_{K_1, K_2} R^{(2)a}_{I, K_1 K_2} \Phi_{K_1} \Phi_{K_2} 
+ \cdots 
\ee
and consider the $m$-th term. 
The corresponding term in eq.\ (\ref{ward2}) becomes 
\ba
\sum_{k=1}^\infty \A\A\!\!\!\!\!\!\! \sum_{i=1}^k \sum_{\{I,K\}} 
j_{I_1} \cdots j_{I_i} \lambda^a R^{(m)a}_{I_i, K_1 \cdots K_m} 
j_{I_{i+1}} \cdots j_{I_k} Z_{I_k, \cdots, I_{i+1}, K_m, 
\cdots, K_1, I_{i-1}, \cdots, I_1} \nonu 
\A = \A \sum_{k=1}^\infty \sum_{\{I, K\}} j_{I_1}  \cdots j_{I_k} 
\lambda^a R^{(m)a}_{I_k, K_1 \cdots K_m} 
{\delta \over \delta j_{K_m}} \cdots 
{\delta \over \delta j_{K_1}} {\delta \over \delta j_{I_{k-1}}} 
\cdots {\delta \over \delta j_{I_1}} Z[j]. 
\ea
Therefore, if we define 
\be
\hat q^a = i \sum_{k=0}^\infty \sum_{\{ I \}, K} 
(-1)^{|a| (|I_1|+ \cdots + |I_k| + |K|)} j_{I_1} \cdots j_{I_k} 
j_K R^a_K ( \mbox{$\delta \over \delta j$} )
{\delta \over \delta j_{I_k}} \cdots {\delta \over \delta j_{I_1}}, 
\label{mastercharge2}
\ee
the Ward-Takahashi identities (\ref{ward2}) can be rewritten as 
\be
\hat q^a Z[j] = 0. 
\label{ward3}
\ee
By the similarity transformation (\ref{sim}) we further define 
the master charges 
\ba
\hat Q^a 
\A = \A  Z[j]^{-1} \hat q^a Z[j] \nonu 
\A = \A \sum_{k=0}^\infty \sum_{\{ I \}} 
(-1)^{|a| (|I_1|+ \cdots + |I_k|)} 
\hat \Pi_{I_1} \cdots \hat \Pi_{I_k} 
\hat S^a \, \hat \Phi_{I_k} \cdots \hat \Phi_{I_1}, 
\label{mastercharge}
\ea
where 
\be
\hat S^a = i \sum_K (-1)^{|a| |K|} \hat \Pi_K R^a_K(\hat \Phi). 
\label{sdef}
\ee
Then the Ward-Takahashi identities (\ref{ward3}) become 
\be
\hat Q^a \ket{\Omega} = 0. 
\label{masterward}
\ee
\par
%
%%%%%  algebra of charges  %%%%%%%%%%%%%%%%%%%%%%%%%%%%%%%%%%%%%%%%
%
To understand a physical meaning of the master charges $\hat Q^a$ 
appearing in the 
Ward-Takahashi identities (\ref{masterward}) let us compute the 
algebra among $\hat Q^a$, $\hat\Phi^a$ and $\hat\Pi^a$. 
Commutation relations between $\hat Q^a$ and $\hat\Phi^a$, 
$\hat\Pi^a$ are easily obtained as 
\ba
i [ \hat Q^a, \, \hat \Phi_I \} 
\A = \A -(-1)^{|a| |I|} \, i \, \hat \Phi_I \hat S^a 
= R_I^a (\hat\Phi), \nonu
i [ \hat Q^a, \, \hat \Pi_I \} \A = \A i \hat S^a \hat \Pi_I. 
\label{mastergenerator}
\ea
We see that the master charges $\hat Q^a$ generate the same 
transformations on the master fields $\hat\Phi_I$ as the quantum 
charges $Q^a$ generate on the quantum fields $\Phi_I$ (\ref{generator}). 
The commutation relation among $\hat Q^a$ is found to be 
\be
[ \hat Q^a, \hat Q^b \} = i f^{abc} \hat Q^c, 
\label{masteralgebra}
\ee
which is the same algebra as that of the quantum 
charges (\ref{algebra}). 
This can be proved as follows. First we compute 
\ba
[ \hat Q^a, \hat Q^b \} 
\A = \A \biggl[ \hat Q^a, \; \sum_{k=0}^\infty \sum_{\{ I \}} 
(-1)^{|b| (|I_1|+ \cdots + |I_k|)} 
\hat \Pi_{I_1} \cdots \hat \Pi_{I_k} 
\hat S^b \hat \Phi_{I_k} \cdots \hat \Phi_{I_1} \biggr\} \nonu 
\A = \A \sum_{k=0}^\infty  \sum_{\{ I \}} 
(-1)^{(|a|+|b|)(|I_1|+\cdots+|I_k|)} 
\biggl( \hat \Pi_{I_1} \cdots \hat \Pi_{I_k} 
[ \hat Q^a, \hat S^b \} 
\hat \Phi_{I_k} \cdots \hat \Phi_{I_1} \nonu 
\A \A + \sum_{i=1}^k (-1)^{|a|(|I_i|+\cdots+|I_k|)} 
\hat\Pi_{I_1} \cdots \hat\Pi_{I_{i-1}} \hat S^a \hat\Pi_{I_i} 
\cdots \hat\Pi_{I_k} \hat S^b \hat\Phi_{I_k} \cdots 
\hat\Phi_{I_1} \nonu 
\A \A - \sum_{i=1}^k (-1)^{|a|(|b|+|I_i|+\cdots+|I_k|)} 
\hat\Pi_{I_1} \cdots \hat\Pi_{I_k} \hat S^b \hat\Phi_{I_k} 
\cdots \hat\Phi_{I_i} \hat S^a \hat\Phi_{I_{i-1}} \cdots 
\hat\Phi_{I_1} \biggr), 
\label{masteralgebra2}
\ea
where $k=0$ terms for the second and the third terms in the last 
equality are defined to be zero. 
We have used the commutation relations (\ref{mastergenerator}). 
By rearranging the summations and using the definition of $\hat Q^a$ 
in eq.\ (\ref{mastercharge}), we obtain 
\ba
[ \hat Q^a, \hat Q^b \} 
\A = \A \sum_{k=0}^\infty \sum_{\{ I \}} 
(-1)^{(|a|+|b|)(|I_1|+ \cdots + |I_k|)} 
\hat \Pi_{I_1} \cdots \hat \Pi_{I_k} \nonu 
\A \A \times \biggl( [ \hat Q^a, \hat S^b \} 
- (-1)^{|a||b|} [ \hat Q^b, \hat S^a \} 
- [ \hat S^a, \hat S^b \} \biggr) 
\hat\Phi_{I_k} \cdots \hat\Phi_{I_1}. 
\label{algebracal1}
\ea
Substituting eq.\ (\ref{sdef}) and using eqs.\ (\ref{mastergenerator}), 
(\ref{jacobi2}), the expression in the brackets in 
eq.\ (\ref{algebracal1}) becomes 
\ba
i \sum_{K} \A\A\!\!\!\!\!\!\! (-1)^{|K|(|a|+|b|)} \hat\Pi_K 
\biggl( [ \hat Q^a, R_K^b(\hat\Phi) \} 
- (-1)^{|a||b|} [ \hat Q^b, R_K^a(\hat\Phi) \} \biggr) \nonu 
\A = \A - f^{abc} \sum_K (-1)^{|K||c|} \, 
\hat\Pi_K R_K^c(\hat\Phi) \nonu 
\A = \A i f^{abc} \hat S^c. 
\label{algebracal2}
\ea
Substituting this result back in eq.\ (\ref{algebracal1}) 
we obtain the commutation relation (\ref{masteralgebra}). 
\par
%
%%%%%  summary  %%%%%%%%%%%%%%%%%%%%%%%%%%%%%%%%%%%%%%%%%%%%
%
To summarize, we have shown that the Ward-Takahashi identities 
in large $N$ field theories can be expressed 
as eq.\ (\ref{masterward}) using the master fields. 
The master charges $\hat Q^a$ satisfy the commutation relations 
(\ref{mastergenerator}) and (\ref{masteralgebra}), and are 
generators of the symmetry transformations realized on the space 
of master fields. 
We should note that the existence of symmetry generators 
for master fields is not an obvious fact. 
It is somewhat surprising that the same algebra can be realized 
on the space of master fields as on the space of quantum operators 
because master field operators just represent $N \rightarrow \infty$ 
limit of $N \times N$ matrices and have nothing to do with quantum 
operators. 
\par
Our formula of the master charges (\ref{mastercharge}) gives, 
as a special case, the spacetime translation generator 
(momentum operator) $\hat P^\mu$ for master fields, 
which was given in eq.\ (5.6) of ref.\ \cite{GG}. 
In this case the operator in eq.\ (\ref{mastercharge2}) commutes 
with $Z(j)$, and therefore coincides with the master charge 
in eq.\ (\ref{mastercharge}). 
\par
In the case of SU($N$) gauge theories the most fundamental symmetry 
is the BRST symmetry. According to the above general result 
we can construct the master charge $\hat Q_{\rm B}$ for the BRST 
symmetry, which generates the BRST transformation of master fields. 
The BRST transformation of master fields was previously discussed 
in ref.\ \cite{AV} in another context. 
%
%\vspace{5mm}
%
%%%%%%%  Acknowledgement  %%%%%%%%%%%%%%%%%%%%%%%%%%%%%%%%%%
%

%
%%%%%%%  References  %%%%%%%%%%%%%%%%%%%%%%%%%%%%%%%%%%%%%%%
\vspace{5mm}
%\newpage
%
\newcommand{\NP}[1]{{\it Nucl.\ Phys.\ }{\bf #1}}
\newcommand{\PL}[1]{{\it Phys.\ Lett.\ }{\bf #1}}
\newcommand{\CMP}[1]{{\it Commun.\ Math.\ Phys.\ }{\bf #1}}
\newcommand{\MPL}[1]{{\it Mod.\ Phys.\ Lett.\ }{\bf #1}}
\newcommand{\IJMP}[1]{{\it Int.\ J. Mod.\ Phys.\ }{\bf #1}}
\newcommand{\PR}[1]{{\it Phys.\ Rev.\ }{\bf #1}}
\newcommand{\PRL}[1]{{\it Phys.\ Rev.\ Lett.\ }{\bf #1}}
\newcommand{\PTP}[1]{{\it Prog.\ Theor.\ Phys.\ }{\bf #1}}
\newcommand{\PTPS}[1]{{\it Prog.\ Theor.\ Phys.\ Suppl.\ }{\bf #1}}
\newcommand{\AP}[1]{{\it Ann.\ Phys.\ }{\bf #1}}
\end{document}